\documentclass[10pt]{iopart}

\usepackage{graphicx}
\usepackage{color}

\begin{document}

\title[Properties of Heusler alloys calculated from
first-principles]{Spin-polarization and electronic properties of
half-metallic Heusler alloys calculated from first-principles}

\author{I Galanakis\dag\ and Ph Mavropoulos\ddag}

\address{\dag\ Department of Materials Science, School of Natural
  Sciences, University of Patras, Patras 265 04, Greece}
\address{\ddag\ Institut f\"ur Festk\"orperforschung, Forschungszentrum J\"ulich, D-52425
J\"ulich, Germany}

 \ead{galanakis@upatras.gr,ph.mavropoulos@fz-juelich.de}

\begin{abstract}
Half-metallic Heusler alloys are amongst the most promising materials
for future magnetoelectronic applications. We review some recent
results on the electronic properties of these compounds. The origin of
the gap in these half-metallic alloys and its connection to the
magnetic properties are well understood. Changing the lattice
parameter shifts slightly the Fermi level. Spin-orbit coupling induces
states within the gap but the alloys keep a very high degree of
spin-polarization at the Fermi level. Small degrees of doping and
disorder as well as defects with low formation energy have little
effect on the properties of the gap, while temperature effects can
lead to a quick loss of half-metallicity. Finally we discuss two
special issues; the case of quaternary Heusler alloys and the
half-metallic ferrimagnets.
\end{abstract}


\submitto{\JPCM}

\maketitle

\section{Half-metallic Heusler Alloys \label{sec1}}

The recent development in electronics, combining the magnetic and
semiconducting materials (so-called magnetoelectronis or spintronics),
has brought half-metallic ferromagnets, initially predicted by de
Groot and collaborators in 1983 \cite{deGroot}, to the centre of
scientific research.  These magnetic metals have the peculiarity that
the minority-spin band-structure is semiconducting while the
majority-spin band-structure is metallic.  Such half-metals exhibit,
ideally, a 100\% spin polarization at the Fermi level and therefore
these compounds should have a fully spin-polarized current and should
be ideal spin injectors into a semiconductor, thus maximizing the
efficiency of spintronic devices~\cite{Zutic2004}.

An important part of the scientific effort has been focused on the
study and fabrication of half-metallic Heusler alloys. The first
family of Heusler alloys studied were of the form X$_2$YZ,
crystallizing in the $L2_1$ structure which consists of four fcc
sublattices, where X a high valent transition or noble metal atom, Y a
low-valent transition metal atom and Z an sp element
\cite{landolt,landolt2}. Such Heusler compounds have attracted a lot
of interest due to the possibility to study in the same family of
alloys a series of interesting diverse magnetic phenomena like
itinerant and localized magnetism, antiferromagnetism, helimagnetism,
Pauli paramagnetism or heavy-fermionic behaviour
\cite{landolt,landolt2}.  The Heusler alloys of the second class are
of the form XYZ, crystallizing in the $C1_b$ structure, and consisting
of three fcc sublattices; they are often called half- or semi-Heusler
alloys in literature, while the $L2_1$ compounds are referred to as
full Heusler alloys. The interest in these types of intermetallic
alloys was revived after the prediction \cite{deGroot}, using
first-principles calculations, of half-metallicity in NiMnSb, a
half-Heusler compound.

The main advantages of Heusler alloys with respect to other
half-metallic systems (e.g. some oxides like CrO$_2$ and
Fe$_3$O$_4$ and some manganites like La$_{0.7}$Sr$_{0.3}$MnO$_3$)
\cite{Soulen98}) are their relatively high Curie temperatures
\cite{landolt,landolt2}. While for the other compounds the Curie
temperature is near the room temperature,  \textit{e.g.} for
NiMnSb it is 730 K and for Co$_2$MnSi it reaches the 985 K
\cite{landolt}. and their structural similarity to the zinc-blende
structure, adopted by binary semiconductors widely used in
industry (such as GaAs on ZnS). Heusler alloys have been already
incorporated in spin-filters \cite{Kilian00}, tunnel junctions
\cite{Tanaka99}, and GMR devices \cite{Caballero98}. The most
successful recent applications in spintronics concern the
half-metallic full Heusler alloys. The group of Westerholt has
incorporated Co$_2$MnGe in the case of spin-valves and multilayer
structures \cite{Westerholt}. The group of Reiss managed to create
magnetic tunnel junctions based on Co$_2$MnSi
\cite{Reiss04,Reiss04b}. A similar study by Sakuraba and
collaborators resulted in the fabrication of magnetic tunnel
junctions using Co$_2$MnSi as magnetic electrodes and Al-O as the
barrier and their results are consistent with the presence of
half-metallicity for Co$_2$MnSi \cite{Sakuraba}. Dong and
collaborators recently managed to inject spin-polarized current
from Co$_2$MnGe into a semiconducting structure \cite{Dong}.

In a recent article \cite{Review-JPD} we have reviewed theoretical
results of the basic electronic and magnetic properties of the
half-metallic Heusler alloys. The gap has its origin for both the
half- and full-Heusler alloys in the $d$-$d$ hybridisation, which is
fundamental for understanding their electronic and magnetic
properties. For both families of compounds the total spin magnetic
moment scales with the number of valence electrons and can be
described by a Slater-Pauling rule.

In the present contribution we review our most recent results on the
electronic properties of these alloys obtained from first-principles
electronic structure calculations. In section \ref{sec2} we summarize
our older results (see references \cite{Review-JPD} and
\cite{Springer} for an extended review). The following sections are
dedicated to phenomena which can destroy half-metallicity. In
particular section \ref{sec3} discuss the effect of the lattice
parameter, section \ref{sec4} is dedicated to the effect of spin-orbit
coupling, section \ref{sec6} to the doping and disorder effects and
the formation of defects. In sections \ref{sec9} and \ref{sec10} we
discuss two special cases: the quaternary Heusler compounds and the
case of half-metallic ferrimagnets, respectively.  In section
\ref{sec11} we review the theoretical studies on the behaviour of the
half-metallicity and of the magnetization as a function of the
temperature. Finally in section \ref{sec12} we summarize and conclude
our review. Note that the problems of surface/interface states
\cite{GalaSurfInterf,Jenkins} are reviewed in other papers in this
volume.

To obtain these results we have employed a variety of electronic
structure methods each one suitable for the property under
investigation. The results in section \ref{sec2}, \ref{sec3} and
\ref{sec4} have been obtained using the full-potential version of the
screened Korringa-Kohn-Rostoker (KKR) method \cite{KKR}. The
relativistic calculations to produce the orbital moments in \ref{sec4}
are performed with the fully-relativistic version of the KKR method
where the Dirac equation is solved \cite{Hubert}.  Results in sections
\ref{sec6} and \ref{sec10} come from the full--potential nonorthogonal
local--orbital minimum--basis band structure scheme (FPLO)
\cite{FPLO}. Finally for the study of the quaternary Heusler alloys
the KKR method is employed within the coherent potential approximation
(CPA) to simulate the disorder in the crystal \cite{KKR-CPA}.

\section{Electronic and gap properties -- Slater Pauling behaviour}
\label{sec2}

The electronic, magnetic and gap properties of half-Heusler alloys
have been extensively studied in reference \cite{GalanakisHalf}
and of full-Heusler compounds in reference \cite{GalanakisFull}.
These results have been extensively also reviewed in reference
\cite{Review-JPD} and in the introductory chapter of
\cite{Springer} and the reader is directed to them for an extended
discussion.  In this section we will only briefly overview these
properties.

Several Heusler alloys have been predicted to be half-metals
\cite{various}; perhaps the most widely studied among them is NiMnSb,
the first to be predicted as a half-metal. The electronic
band-structure calculations of NiMnSb reveal that for both spin
directions there is one band at around $-12$ eV with reference to the
Fermi level which arises from the $s$ states of the Sb atom. The next
three lowest bands are due to the $p$ states of Sb and accommodate
also electrons from the transition metal atoms
\cite{GalanakisHalf}. The Ni and Mn majority-spin $d$ states are
energetically close and form a common band. The situation is different
for the minority-spin states: due to the strong exchange splitting of
Mn, the minority Mn $d$-states are unoccupied, while the ones of Ni
are occupied. The minority-spin gap arises between the bonding and
antibonding $d$-hybrids created among the Mn and Ni states (in this
respect, the origin of the gap is similar to the one in compound
semiconductors like GaAs or ZnSe). The minority-spin bonding states
have most of their weight at the Ni atom and are occupied, while the
antibonding states have their weight mostly at the Mn atom and are
unoccupied. This leads to very strong localized spin moments at the Mn
atoms (see table \ref{tableorb} for the values of the moments)
\cite{Plogmann}.  A simple band counting shows that the total spin
moment is exactly 4 $\mu_\mathrm{B}$ for NiMnSb; this is a secial case
of the Slater-Pauling behaviour, leading to integer moments for
half-metallic systems.

The total moment of the half-metallic $C1_b$ Heusler alloys
follows a simple rule. The total number of electrons, $Z_t$, is
given by the sum of the number of spin-up and spin-down electrons,
while the total moment $M_t$ per unit cell is given by the
difference. Since 9 minority bands are fully occupied, we obtain
the simple ``rule of 18'' for half-metallicity in $C1_b$ Heusler
alloys \cite{GalanakisHalf,jung}
\begin{equation}
M_t = Z_t - 18.\label{eq:SP1}
\end{equation}
This behaviour of the total moment resembles the Slater-Pauling
behaviour of the binary transition metal alloys with the
difference that in the latter ones the total spin moment decreases
with the number of valence electrons \cite{Kubler}. The
half-Heusler alloys with 18 valence electrons (e.g., CoTiSb) are
nonmagnetic semiconductors, where the gap is again created by the
$d$-$d$ hybridization.

\begin{figure}
\centering
\includegraphics[width=0.45\linewidth]{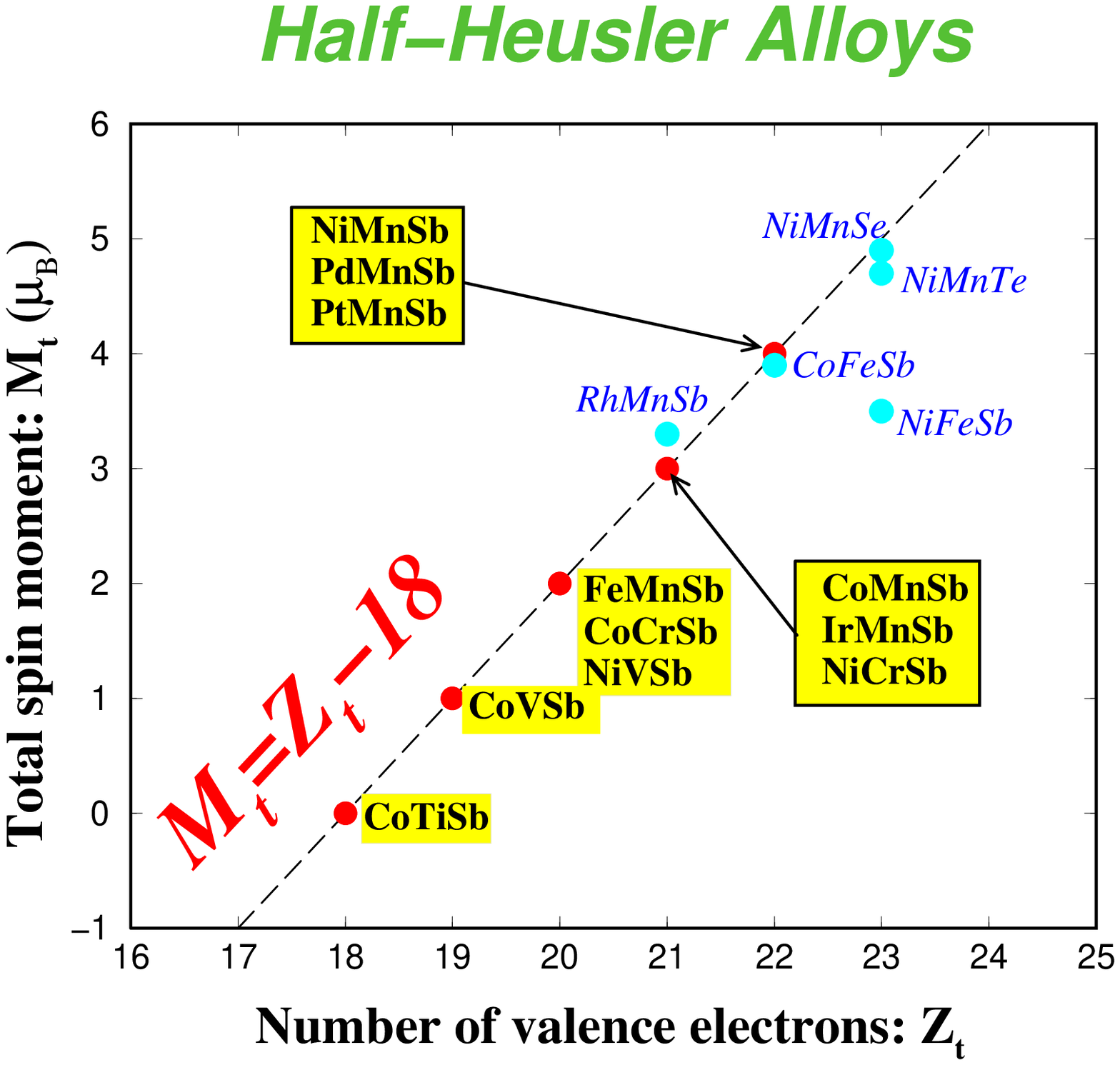}
\includegraphics[width=0.45\linewidth]{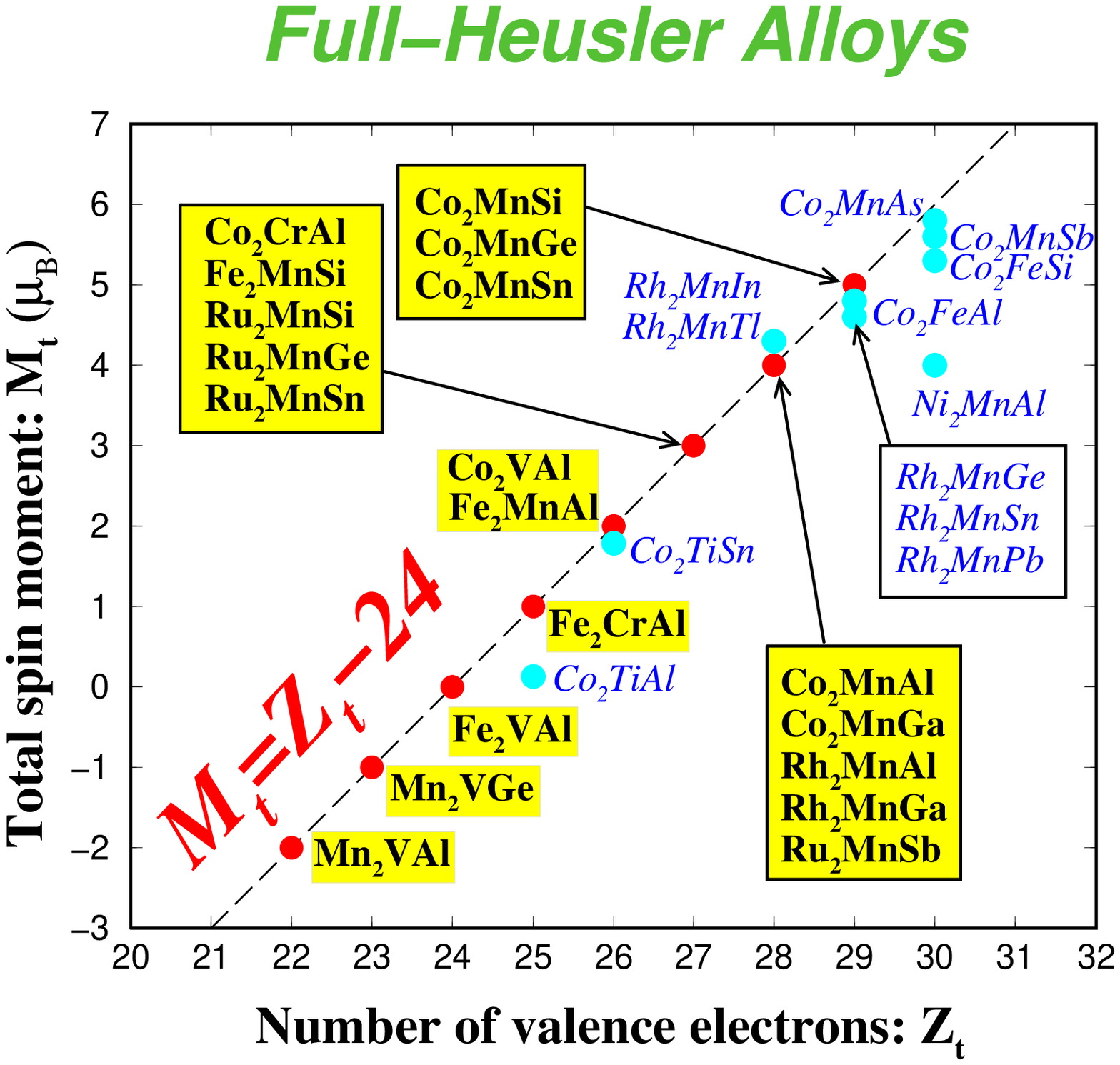}
\caption{(Color online) Calculated total spin moment per unit cell
as a function of  the total number $Z_t$ of valence electrons per
unit cell for all  the studied half (left panel) and full (right
panel) Heusler alloys. The dashed line represents the
Slater-Pauling behaviour.}
 \label{fig1}
\end{figure}

In figure \ref{fig1} (left panel) we have gathered the calculated spin
magnetic moments per formula unit for the half-Heusler alloys which we
have plotted as a function of the total number of valence
electrons. The dashed line represents the rule $ M_t = Z_t -18$ obeyed
by these compounds. The total moment $M_t$ is an integer (in $\mu_B$),
assuming the values 0, 1, 2, 3, 4 and 5 if $Z_t \ge 18$.  The value
$M_t=0$ corresponds to the semiconducting phase and the value $M_t=5\
\mu_B$ to the maximal moment when all 5 majority $d$-states are
filled.  Firstly we varied the valence of the lower-valent
(\textit{i.e.}  magnetic) transition metal atom. Thus we substitute V,
Cr and Fe for Mn in the NiMnSb and CoMnSb compounds using the
experimental lattice constants. For all these compounds we find that
the total spin moment scales accurately with the total charge and that
they all present half-metallicity.

In the second part of this section we discuss the properties of the
full-Heusler alloys like Co$_2$MnSi and Co$_2$CrAl. Their electronic
and magnetic properties are similar to the half-Heusler compounds with
the additional complication of the presence of 2 Co atoms per unit
cell. In addition to the bands the half-Heusler alloys, there are five
states exclusively located at the Co sites and near the Fermi level
\cite{GalanakisFull}. Now, the Fermi level is located among these
states so that three out of five are occupied and two of them
unoccupied leading to smaller gaps. Now there are 12 minority-spin
occupied states and the compounds with 24 valence electrons like
Fe$_2$VSb are semiconductors. If V is substituted by a higher-valent
atom, spontaneous spin polarization occurs, and the exchange splitting
shifts the majority states to lower energies. Thus the extra electrons
fill in only majority states: Co$_2$CrAl (27 valence electrons) has a
total spin moment of 3 $\mu_\mathrm{B}$ and Co$_2$MnSi (29 valence
electrons) a spin moment of 5 $\mu_\mathrm{B}$ as can be seen in table
\ref{tableorb}.

The presence of the second Co atom has also another important
consequence. Now each Mn atom has an octehedral environment of Co
atoms as first neighbours and not four as was the case for CoMnSb.
This leads to an increased hybridization between the $d$ orbitals of
the Mn and Co atoms and the spin moment of Co in Co$_2$MnSi is
slightly lower than 1 $\mu_\mathrm{B}$, considerably larger than the
Co moment in CoMnSb.

Following the above discussion we investigate the Slater-Pauling
behaviour with the ``rule of 24'' now instead of 18 in the
half-Heuslers :
\begin{equation}
M_t = Z_t - 24.
\label{eq:SP2}
\end{equation}
In figure \ref{fig1} (right panel) we have plotted the total spin
moments for all the studied compounds as a function of the total
number of valence electrons. The dashed line represents the
Slater-Pauling rule of half-metallic full Heusler alloys. The
``magical number 24'' arises from the fact that the minority band
contains 12 electrons per unit cell: 4 are occupying the low lying $s$
and $p$ bands of the $sp$ element and 8 the Co-like minority $d$ bands
(the Mn-Co bonding hybrids: $2\times e_g$ and $3\times t_{2g}$, and
the non-bonding Co states localized only at the Co sites: $3\times
t_{1u}$). Since 7 minority bands (2$\times$Co $e_u$, 5$\times$Mn $d$)
are unoccupied, the largest possible moment is 7 $\mu_B$ and would
occur if all majority $d$-states were occupied. Of course it would be
impossible to get a compound with a total spin moment of 7 $\mu_B$ but
even $M_t=6\ \mu_B$ is difficult to obtain. As it was shown by Wurmehl
et al.~\cite{Co2FeSi}, the on-site correlations in Co$_2$FeSi play a
critical role for this compound and calculations within the LDA+U
scheme, rather than the LDA, give a spin moment of 6 $\mu_B$.

The exchange interactions, stabilizing ferromagnetism, have been
studied in reference \cite{GalanakisExchConst} using the adiabatic
(frozen-magnon) approximation to calculate the inter-atomic exchange
parameters for the half-Heusler alloys NiMnSb and CoMnSb. The dominant
interaction is between the Mn atoms. The magnetic interactions are
more complex in full-Heusler alloys Co$_2$MnSi and Co$_2$CrAl
\cite{GalanakisExchConst,Kurtulus}. In both cases, ferromagnetism is
stabilized by the inter-sublattice interactions between the Mn(Cr) and
Co atoms and between Co atoms belonging to different sublattices.

Before closing this section we discuss also the role of the
$sp$-elements in half-metallic heusler alloys.  While the
$sp$-elements are not responsible for the appearance of the minority
gap, they are nevertheless very important for the physical properties
of the Heusler alloys and the structural stability of the $C1_b$
structure. There are three important features:

(i) While an Sb atom has 5 valence electrons (5$s^2$, 5$p^3$), in
the NiMnSb compound each Sb atom introduces a deep lying $s$-band,
at about $-12$ eV, and three $p$-bands below the centre of the
$d$-bands. These bands accommodate a total of 8 electrons per unit
cell, so that formally Sb acts as a triply charged Sb$^{3-}$ ion.
Analogously, a Te-atom behaves in these compounds as a Te$^{2-}$
ion and a Sn-atom as a Sn$^{4-}$ ion. This does not mean, that
locally such a large charge transfer exists. In fact, the $s$- and
$p$-states strongly hybridize with the transition metal (TM)
$d$-states and the charge in these bands is delocalized and
locally Sb even loses about one electron, if one counts the charge
in the Wigner-Seitz cells. What matters here is that the $s$- and
$p$-bands accommodate 8 electrons per unit cell, thus  reducing
the electrons which have to be accommodated by the $d$-bands of
the TM atoms.

(ii) The $sp$-atom is very important for the structural stability of
the Heusler alloys. For instance, it is difficult to imagine that the
calculated half-metallic NiMn and PtMn alloys with zinc-blende
structure actually exist, since metallic alloys prefer highly
coordinated structures like fcc, bcc, hcp etc.  Therefore the
$sp$-elements are decisive for the stability of the $C1_b$
compounds. A careful discussion of the bonding in these compounds has
been recently published by Nanda and Dasgupta \cite{nanda-dasgupta}
using the crystal orbital Hamiltonian population (COHP) method. For
the semiconductor FeVSb they find that, while the largest contribution
to the bonding arises from the V-$d$ -- Fe-$d$ hybridization,
contributions of similar size arise also from the Fe-$d$ -- Sb-$p$ and
the V-$d$ -- Sb-$p$ hybridization. Similar results are also valid for
the semiconductors CoTiSb and NiTiSn and in particular for the
half-metal NiMnSb. Since the majority $d$-band is completely filled,
the major part of the bonding arises from the minority band, so that
similar arguments as for the semiconductors apply.

(iii) Another property of the $sp$-elements is worthwhile
mentioning: substituting the Sb atom in NiMnSb by Sn, In or Te
destroys the half-metallicity \cite{GalanakisHalf}. This is in
contrast to the substitution of Ni by Co or Fe, which is
documented in table \ref{tableorb}. The total moment of 4 $\mu_B$
for NiMnSb is reduced to 3 $\mu_B$ in CoMnSb and 2 $\mu_B$ in
FeMnSb, thus preserving half-metallicity. In NiMnSn the total
moment is reduced only to 3.3 $\mu_B$ (instead of 3) and in NiMnTe
the total moment increases only to 4.7 $\mu_B$ (instead of 5).
Thus by changing only the $sp$-element it is rather difficult to
preserve the half-metallicity, since the density of states changes
more like in a rigid band model \cite{GalanakisHalf}.

\section{Effect of the lattice parameter}\label{sec3}

\begin{figure}
\centering
\includegraphics[scale=0.4]{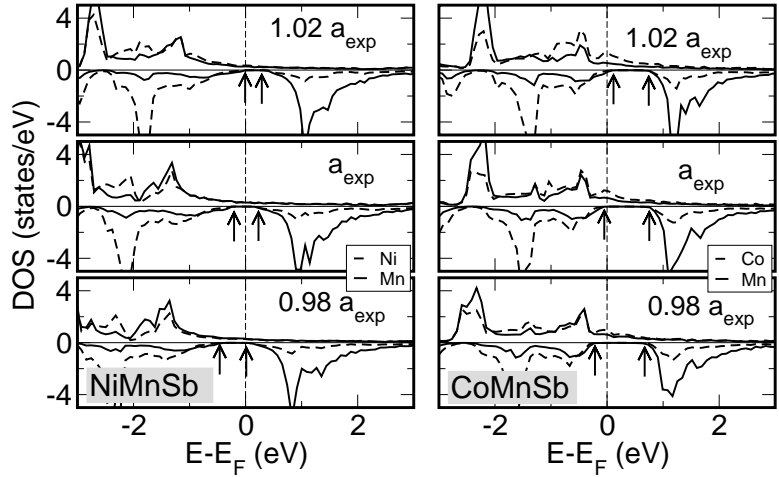}
\caption{Atom-resolved density of states (DOS) for the
experimental lattice parameter for NiMnSb and CoMnSb, compared
with the once compressed or expanded by 2\%. Arrows indicate the
edges of the minority gap.} \label{fig2}
\end{figure}

Changes of the lattice parameter can influence the electronic and
magnetic properties of the $C1_b$ and $L2_1$ Heusler alloys. To
the best of our knowledge no relevant experimental study exists.
In figure \ref{fig2} we show the DOS of NiMnSb and CoMnSb for the
experimental lattice parameter and the ones compressed and
expanded by 2 \%. First one sees that, upon compression, the Fermi
level moves in the direction of the conduction band, upon
expansion towards the valence band. In both cases, however, the
half-metallic character is conserved. In order to explain this
behaviour, we first note that the position of Fermi level is
determined by the metallic DOS in the majority band. As we
believe, the shift of $E_F$ is determined by the behaviour of the
Sb $p$-states, in particular by the large extension of these
states as compared to the $d$ states. Upon compression the
$p$-states are squeezed and hybridize stronger, thus pushing the
$d$-states and the Fermi level to higher energies, i.e., towards
the minority conduction band. In addition the Mn $d$ and Ni or Co
$d$ states hybridize stronger, which tends to increase the size of
the gap while the Mn(Ni) $d$-bandwidth increases, which tends to
shrink the gap. As shown in figure \ref{fig2} the first of the two
effects is stronger leading finally to an increase of the gap
width. Upon expansion the opposite effects are observed. In the
case of NiMnSb and for the experimental lattice constant the
gap-width is $\sim$ 0.4 eV. When the lattice is expanded by 2\%\
the gap shrinks to 0.25 eV and when compressed by 2\%\ the
gap-width is increased to 0.5 eV. Similarly in the case of CoMnSb,
the gap is 0.8 eV for the experimental lattice constant, 0.65~eV
for the 2\%\ expansion and 0.9 eV for the case of the 2\%\
compression.

For the full-Heusler alloys the pressure dependence has been recently
studied by Picozzi et al.~\cite{picozzi} for Co$_2$MnSi, Co$_2$MnGe
and Co$_2$MnSn, using both the LDA and the somewhat more accurate
GGA. The general trends are similar: the minority gap increases with
compression, and the Fermi level moves in the direction of the
conduction band. For example in the case of Co$_2$MnSi the gap-width
is 0.81 eV for the theoretical equilibrium lattice constant of 10.565
\AA . When the lattice constant is compressed to $\sim$ 10.15 \AA, the
gap-width increases to about 1 eV.  The calculations show that for the
considered changes of the lattice constant of $\pm$ 2 \%,
half-metallicity is preserved. There can be sizeable changes of the
local moments, but the total moment remains constant, since $E_F$
stays within the gap.

The results presented in this section up to now concern the
homogeneous change of the lattice constant under hydrostatic
pressure. Block and collaborators have studied the effect of
tetragonalization on the properties of both NiMnSb and Co$_2$CrAl
\cite{Block}. They found that although tetragonalization largely
affected the shape and width of the gap, the systems remained
half-metallic for moderate degrees of strain.

\begin{table}
\caption{\label{tableorb} Calculated spin ($m_\mathrm{spin}$) and
orbital ($m_\mathrm{orb}$) magnetic moments in $\mu_B$ for the
XMnSb half-Heusler and X$_2$YZ full-Heusler compounds. The last
three columns are the total spin and orbital magnetic moment and
their sum, respectively. The indexes X, Y, Z refer to the
corresponding atoms in the compound.}
\begin{tabular}{rrrrrrrrrr}
\hline Alloy& $m^{\mathrm{X}}_\mathrm{spin}$ & $m^{\mathrm{X}}_\mathrm{orb}$ &
$m^{\mathrm{Y}}_\mathrm{spin}$ & $m^{\mathrm{Y}}_\mathrm{orb}$ & $m^{\mathrm{Z}}_\mathrm{spin}$ &
$m^{\mathrm{Z}}_\mathrm{orb}$ & $m^{{\mathrm{total}}}_\mathrm{spin}$ &
$m^{{\mathrm{total}}}_\mathrm{{\mathrm{orb}}}$ & $m^{{\mathrm{total}}}$
\\ \hline FeMnSb  & -0.97&  -0.06&  2.94 &  0.03& -0.04 & $\sim$
-0&  1.96  & -0.03& 1.93\\ CoMnSb  &-0.16&
-0.04& 3.20& 0.03& -0.10& $\sim$ -0  &2.96  & -0.01  &2.95 \\
NiMnSb &0.25& 0.02 &3.72& 0.03  & -0.07&  $\sim$ -0  & 3.95 &
0.04 &3.99 \\
\hline Co$_2$MnAl& 0.75 &  0.01 &  2.60 &0.01& -0.09 & $\sim$ 0 &
4.00 & 0.04 &4.04
\\ Co$_2$MnSi & 0.99 &0.03& 3.02& 0.02 &  -0.08 & $\sim$ 0 &
4.93& 0.08 &  5.00 \\  Co$_2$CrAl& 0.70& 0.01 & 1.64 & 0.01
&-0.08&  $\sim$ 0 &2.97 &  0.03& 3.00
\\  Mn$_2$VAl &
-1.40
&-0.03 &0.79& -0.01& 0.01& 0.01& -2.00& -0.07 &-2.07 \\
\hline
\end{tabular}
\end{table}

\section{Spin-orbit coupling and orbital magnetism}\label{sec4}

The calculations presented up to this point have neglected the
spin-orbit coupling (SOC). Intuitively, however, one expects that SOC
can be of crucial importance for the half-metallic property: In the
presence of SOC, the electron spin is no more a good quantum number,
so that the electron eigenfunctions cannot conserve their spin degree
of freedom. Spin-up wavefunctions within the half-metallic gap must
then have partly spin-down character. As a result, the celebrated
half-metallic gap cannot really be 100\% there even at
$T\rightarrow0$. In materials where the SOC is weak, the DOS within
the ``gap'' is expected to be low, and the polarization close to, but
not exactly at, 100\% \cite{Mavropoulos}. Another result of SOC is the
appearance of orbital magnetic moment. This, too, is weak in Heusler
alloys with low SOC strength.

The spin polarization $P(E)$ at an energy $E$ (and in particular
at $E_F$) is related to the spin-dependent DOS via the expression
\begin{eqnarray}
P&=&\frac{n_{\uparrow}(E_F)-n_{\downarrow}(E_F)}
{n_{\uparrow}(E_F)+n_{\downarrow}(E_F)} \\
&\approx& 1-2\, n_{\downarrow}(E_F)/n_{\uparrow}(E_F) \ \ \
\mbox{for small  $n_{\downarrow}/n_{\uparrow}$ }. \nonumber
\end{eqnarray}

The spin-orbit coupling connecting the two spin channels reads, in
terms of the Pauli matrices $\vec{\sigma}$ and the orbital momentum
$\vec{L}$:
\begin{equation}
V_{\mathrm{so}}(r)=\frac{1}{2m^2c^2}\frac{\hbar}{2}
\frac{1}{r}\frac{dV}{dr}\,\vec{L}\cdot\vec{\sigma} = \left(
\begin{tabular}{cc}
$V_{\mathrm{so}}^{\uparrow\uparrow}$   & $V_{\mathrm{so}}^{\uparrow\downarrow}$ \\
$V_{\mathrm{so}}^{\downarrow\uparrow}$ &
$V_{\mathrm{so}}^{\downarrow\downarrow}$
\end{tabular}
\right). \label{eq:soc1.0}
\end{equation}
Here, $V(r)$ is the unperturbed one-electron potential at an atomic
site, and is assumed to be spherically symmetric.  Deviations from the
spherical symmetry arise only close to the interstitial region between
atoms, where the contribution to the spin-orbit interaction is anyhow
small. The $2\times2$ matrix is the perturbation expressed in spinor
basis, demonstrating the non-diagonal terms
$V_{\mathrm{so}}^{\uparrow\downarrow}$ and
$V_{\mathrm{so}}^{\downarrow\uparrow}$ which are responsible for
spin-flip processes; $\uparrow$ and $\downarrow$ denote the spin up
and spin down direction. Within perturbation theory one can show that
the spin-down DOS within the gap is a weak reflection of the spin-up
DOS, depending quadratically on the spin-orbit coupling strength:
$n_{\downarrow}\sim n_{\uparrow}
(V_{\mathrm{so}}^{\uparrow\downarrow})^2$ \cite{Mavropoulos}. Moreover
it can be shown that, close to the gap edges, there is an enhancement
of the spin-flip elements.

In reference~\cite{Mavropoulos}, $P(E_F)$ was calculated for a number
of half-Heusler alloys and other half-metals. The calculations were
done within density-functional theory, utilising the KKR Green
function method~\cite{Hubert}. The approach was fully relativistic,
solving the Dirac equation rather than treating the spin-orbit
coupling as a perturbation. Nevertheless the qualitative behaviour
which was described above was revealed.  Table~\ref{table1soc}
summarizes the results for $P(E_F)$ and for the polarization in the
middle of the gap, $P(E_M)$, for some half-Heusler alloys. The trend
is that alloys which include heavier elements show a lower spin
polarization. This becomes most evident by inspection of the values of
$P(E_M)$, since for some of the alloys the Fermi level approaches or
enters the valence band, so that $P(E_F)$ does not always reflect the
SOC strength. In particular, NiMnSb shows a high value for $P(E_M)$
(99.3\%), which decreases when we substitute the $3d$ element Ni with
the heavier, $4d$ element, Pd, and even more so when we substitute it
with the even heavier, $5d$ element, Pt. This effect is only expected,
since it is well-known that heavier elements show in general a
stronger SOC due to the term $dV/dr$ in Eq.~\ref{eq:soc1.0}.  As a
conclusion, the spin-orbit coupling reduces $P(E_F)$, but the
resulting values are still high. Nevertheless, we point out that
calculations on half-metallic ferromagnets containing heavy elements
({\it e.g.}, lanthanides) should take into account the spin-orbit
coupling for a reliable quantitative result.

\begin{table}
\begin{center}
\caption{Calculated spin polarisation $P$ at the Fermi level
$(E_F)$
  and in the middle of the spin-down gap $(E_M)$, and ratio of
  spin-down/spin-up DOS in the middle of the gap
  $(n_{\downarrow}/n_{\uparrow})(E_M)$, for various half-Heusler
  alloys. The alloys PdMnSb and PtMnSb present a spin-down gap, but
  are not half-metallic, as $E_F$ is slightly below the
  gap. \label{table1soc}}
\begin{tabular}{ccccc}
\hline
Compound & $P(E_F)$ & $P(E_M)$ & $(n_{\downarrow}/n_{\uparrow})(E_M)$ \\
\hline
CoMnSb   &  99.0\%  &   99.5\% & 0.25\% \\
FeMnSb   &  99.3\%  &   99.4\% & 0.30\% \\
NiMnSb   &  99.3\%  &   99.3\% & 0.35\% \\
PdMnSb   &  40.0\%  &   98.5\% & 0.75\% \\
PtMnSb   &  66.5\%  &   94.5\% & 2.70\% \\
\hline
\end{tabular}
\end{center}
\end{table}

The orbital moments in Heusler alloys are expected to be small.
This is a result of the cubic symmetry, of the fact that the
magnetic transition elements here are relatively light ($3d$
series) and of the metallic nature of the electronic states.

Orbital moments were calculated by Galanakis et al. for half-Heusler
compounds \cite{iosif}, by Picozzi {\it et al.}~\cite{picozzi} for
Co$_2$MnSi, -Ge, and -Sn, and more systematically by
Galanakis~\cite{GalanakisOrbit}. Some of the latter results are
summarized in table~\ref{tableorb}. As expected, the values of the
orbital moments $m_{\mathrm{orb}}$ (calculated within the LSDA) are
small. It is known that the LSDA can underestimate $m_{\mathrm{orb}}$
by up to 50\%, but the trends are considered reliable. The highest
orbital moment, almost 0.1~$\mu_B$, appears at the Ir and Mn atoms in
IrMnSb, but with opposite signs for the two atoms, so that the total
$m_{\mathrm{orb}}$ is close to zero (note that also the Ir spin moment
is opposite to the Mn spin moment; for Ir in this compound we obtain a
ratio of $m_{\mathrm{orb}}/m_{\mathrm{spin}}\simeq 1/2$).

\section{Doping, disorder and defects}\label{sec6}

Before starting our discussion on doping, disorder and defects in
Heusler alloys, we should note that, as shown by Larson et al
\cite{Larson}, the ordered structure of Heusler alloys is the one with
the minimum energy and thus the more stable.

 Disorder in NiMnSb had been studied initially by Orgassa and
collaborators \cite{Orgassa}.  They took into account three
different types of disorder: i) intermixing of Ni and Mn atoms,
ii) migration of Ni and Mn atoms at the empty site (the one
occupied by the second high-valent transition metal atom in the
full Heusler alloys), and iii) migration of Mn and Sb atom to the
empty site. They found that 5\% of disorder was enough so that the
impurity states destroy the minority-spin gap. This work was
recently extended by Alling et al.~\cite{Alling} covering all
possible kinds of interstitial, vacancy and atomic-swap defects in
NiMnSb in the diluted limit. They have shown that defects span a
large area of energy formation values, from 0.2 to 14.4 eV. More
interestingly defects with formation energies less than 4 eV,
which are also the most likely to occur, do not destroy the
half-metallicity although they severely affect the width and shape
of the gap.

The formation of defects in full Heusler alloys and more precisely
in Co$_2$MnGe and Co$_2$MnSi has been studied by Picozzi et al
\cite{PicozziDef}. They found that the Mn antisite have the lowest
formation energy and they retain half-metallicity contrary to the
Co ones. The Mn-Si and Mn-Co atomic swaps were found to have very
large formation energies and thus are unlikely to occur. But these
results cannot be generalized to all Heusler alloys. As shown by
Miura et al in the case of Co$_2$CrAl the Cr-Al swaps have low
formation energy and the system prefers energetically the $B2$
structure where the Cr and Al atoms are completely disordered
\cite{ShiraiQuart}. Substituting Fe for Cr restores the order
since the Fe-Al atomic swaps have very high formation energy and
are unlikely to occur.

Galanakis et al studied the effect of doping and disorder on the
properties of the half-metallic full Heusler alloys: Co$_2$MnSi,
Co$_2$MnGe and Co$_2$MnSn \cite{GalanakisDisorder}.  To simulate
the doping by electrons we substituted Mn by Fe while to simulate
the doping of the alloys with holes we substituted Mn by Cr. We
studied substitution of 5\%, 10\% and 20\% of the Mn atoms. Doping
the perfectly ordered alloys with either Fe or Cr induces a
smoothening of the DOS. Cr-doping has only marginal effects to the
gap. Its width is narrower with respect to the perfect compounds
but overall the compounds retain their half-metallicity. In the
case of Fe-doping the situation is more complex. Adding electrons
to the system means that, in order to retain the perfect
half-metallicity, these electrons should occupy majority
antibonding levels at higher energies. This is energetically not
very favourable and in the case of Co$_2$MnSi and for 20\% Fe
doping the Fermi level is no more exactly in the gap but slightly
above it. In general the total spin moment follows the
Slater-Pauling behaviour with only slight deviations for the
Co$_2$MnSi compound with for Fe-doping.

Finally we discuss the effect of disorder between the Mn and the
$sp$ atoms in Co$_2$MnSi, Co$_2$MnGe and Co$_2$MnSn
\cite{GalanakisDisorder}. If one inspects the atom-resolved DOS,
the gap showing at the DOS of the Mn and $sp$ atoms is wider than
the gap showing at the DOS of the Co atoms. This is because the
states around the ``total'' gap are of Co-character only, as
discussed in section \ref{sec2}. Intermixing Mn and the $sp$ atoms
changes the symmetry of the Co sites and in this way new states in
the gap can be induced, affecting the half-metallic character.
Substituting Mn by Si in Co$_2$MnSi induces states just at the
high edge of the gap while substituting Si by Mn pushes the
unoccupied minority states even higher in energy and the gap
becomes wider. Substituting 5\%, 10\% or 20\% of the Mn atoms by
Si results in a decrease of 0.15, 0.30 and 0.60 of the total
number of valence electrons in the cell, while the inverse
procedure results to a similar increase of the mean value of the
number of valence electrons. The situation is analogous in
Co$_2$MnGe and Co$_2$MnSn.  The compounds containing Si and Ge
show a perfect Slater-Pauling behaviour (as a function of the
substitution fraction $x$) while the Co$_2$Mn$_{1+x}$Sn$_{1-x}$
deviate from the ideal values of the total spin moment.

\section{Quaternary Heusler alloys}\label{sec9}

We proceed with a discussion on the quaternary Heusler alloys
\cite{ShiraiQuart,GalanakisQuart}.  In the these compounds, one of the
four sites is occupied by two different kinds of neighbouring elements,
like Co$_2$[Cr$_{1-x}$Mn$_{x}$]Al, where the Y site is occupied by Cr
or Mn atoms. For all calculations we assumed that the lattice constant
varies linearly with the concentration $x$ which has been verified for
several quaternary alloys \cite{landolt,landolt2}.

\begin{figure}
\centering
\includegraphics[scale=0.5]{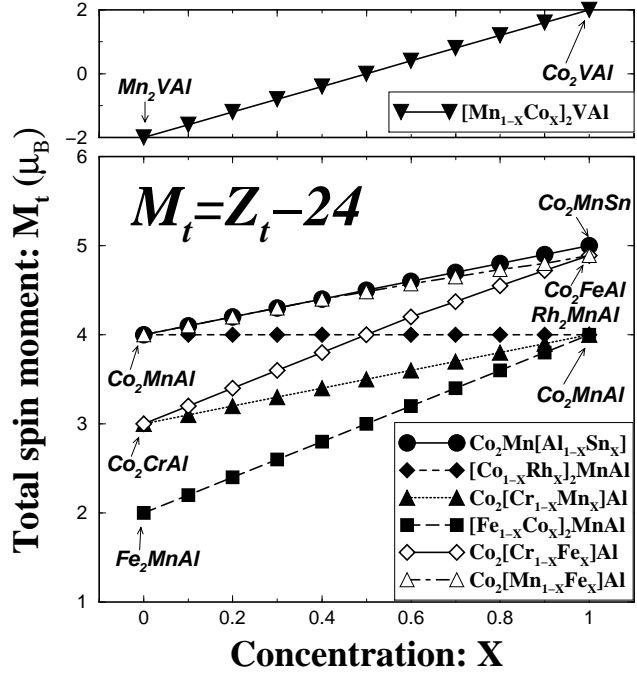}
\caption{Calculated total spin moment $M_t$ in $\mu_B$ for a
variety of compounds as a function of the concentration $x$
($x$=0,0.1,0.2,...,0.9,1). We assumed that the lattice constant
varies linearly with the concentration $x$. With solid lines the
cases obeying the rule $M_t=Z_t-24$ are shown where $Z_t$ and
$M_t$ are the average total number of valence electrons and the
average total moment.} \label{figquart}
\end{figure}

In reference \cite{GalanakisQuart}, calculations are reported on
several quaternary alloys taking into account several possible
combinations of chemical elements and considering a concentration
increment of 0.1. The results are resumed in figure \ref{figquart}. In
the first case, two different low-valent transition metal atoms can
occupy the Y site, as in Co$_2$[Cr$_{1-x}$Mn$_x$]Al. The total spin
moment varies linearly between the 3 $\mu_B$ of Co$_2$CrAl and the 4
$\mu_B$ of Co$_2$MnAl. In the case of Co$_2$[Cr$_{1-x}$Fe$_x$]Al and
Co$_2$[Mn$_{1-x}$Fe$_x$]Al, and up to around $x$=0.6 the total spin
moment shows a Slater-Pauling behaviour but for larger concentrations
$M_t$ deviates slightly; Co$_2$FeAl, corresponding to $x=1$, has a
non-integer moment. Similar results were also obtained in reference
\cite{Antonov}. This behaviour is clearly seen in figure
\ref{figquart} when we compare the lines for the
Co$_2$[Mn$_{1-x}$Fe$_x$]Al and Co$_2$Mn[Al$_{1-x}$Sn$_x$] compounds;
the latter family follows the SP behaviour.
The latter case brings us to the set of compounds where $sp$ elements
are mixed, and these compounds also obey the rule for the total spin
moments.

The third case is to mix the higher valent transition metal atoms, as
in [Fe$_{1-x}$Co$_x$]$_2$MnAl and [Rh$_{1-x}$Co$_x$]$_2$MnAl alloys.
In [Fe$_{1-x}$Co$_x$]$_2$MnAl the total spin moment varies linearly
between the 2 $\mu_B$ of Fe$_2$MnAl and 4 $\mu_B$ of Co$_2$MnAl. As
for [Rh$_{1-x}$Co$_x$]$_2$MnAl, we observe that Rh is isoelectronic to
Co, thus we find a constant integer value of 4 $\mu_B$ for all
concentrations.

Finally, we discuss the special case of Mn$_2$VAl, which has less
than 24 electrons and a total spin moment of $-2$ $\mu_B$. Here,
mixing Mn and Co gives a family of compounds
(Mn$_{1-x}$Co$_x$)$_2$VAl, where the total spin moment varies
linearly between $-2$ $\mu_B$ and 2 $\mu_B$. Interestingly, for
$x$=0.5 we get the case of a compound without macroscopic
magnetization, consisting of magnetic atoms (as in the case of an
antiferromagnet). Thus all the compounds obey the rule
$M_t$=$Z_t-24$, showing the Slater-Pauling behaviour regardless of
the origin of the extra charge.

As a rule of thumb, we expect that for two half-metallic alloys like
XYZ and X$'$YZ (or XY$'$Z or XYZ$'$), which both lie on the
Slater-Pauling curve, also the mixtures like X$_{1-x}$X$'_x$YZ lie on
the Slater Pauling curve, with an average moment of $\left<M_t\right>
= (1-x) M^{\mathrm{XYZ}}_t + x M_t^{\mathrm{X}'\mathrm{YZ}}$. However,
it is not guaranteed that these intermediate structures are stable,
especially if the parent compounds are not neighbours on the
Slater-Pauling curve.

\section{Half-metallic ferrimagnets}\label{sec10}

\begin{table}
\caption{Half metallic lattice parameters  and spin magnetic
moments (in  $\mu_B$) of Mn$_2$VZ (Z=Al, Si, Ge, Sn ). }
\begin{indented}
 \item[]
 \begin{tabular}{lccccc}
  \hline \hline
   Compound& $a(\AA)$ &  $m_\mathrm{Mn}$ & $m_\mathrm{V}$ &
$m_\mathrm{Z}$ & $m_\mathrm{Cell}$  \\
  \hline
Mn$_2$VAl   &5.987& -1.510  & 0.967 & 0.053    & -2.00 \\
            &6.117& -1.655  & 1.233 & 0.076    & -2.00 \\
 \hline
Mn$_2$VSi   &6.06 & -0.863  & 0.675 & 0.052    & -1.00 \\
            &6.29 & -1.092  & 1.105 & 0.078    & -1.00 \\
 \hline
Mn$_2$VGe   &6.18 & -0.976  & 0.905 & 0.048    & -1.00 \\
            &6.27 & -1.082  & 1.111 & 0.054    & -1.00 \\
 \hline
Mn$_2$VSn   &6.25 & -0.980  & 0.922 & 0.038    & -1.00 \\
            &6.31 & -1.050  & 1.059 & 0.041    & -1.00 \\
  \hline \hline
\end{tabular}
\end{indented}
 \label{tableHFerri}
\end{table}

The ideal case for realistic applications would be a half-metallic
antiferromagnet rather than the half-metallic ferromagnets. In
such a compound the majority- and minority-spin densities of
states are not connected by a symmetry transformation, as for
common antiferromagnets and the material is better described as a
fully compensated ferrimagnet, having a net magnetic moment that
is, due to the half-metallic character, precisely equal to zero.
Such a material would have the advantage that due to the absence
of a total magnetic moment it would not give rise to stray fields.
Van Leuken and de Groot have suggested a possible route towards a
half-metallic antiferromagnet starting from the semiconducting
$C1_b$-type compound FeVSb \cite{Leuken}.  Due to the difficulties
to construct such a material, it would be useful to study also the
half-metallic ferrimagnetic materials, like FeMnSb \cite{FeMnSb}
or the Mn$_2$VZ \cite{Weht} compounds where the Mn and Fe or V
spin moments are anti-parallel.

In reference \cite{GalanakisHFerri} we studied the possibility of
appearance of half-metallicity in the case of the full-Heusler
compounds Mn$_2$VZ where Z is an $sp$ atom belonging to the IIIB or
IVB column of the periodic table. Total energy calculations have shown
that, when Z is Al, Ga, In or Sn, the compounds are ferrimagnetic,
while the compounds containing Si and Ge are non-magnetic at the
equilibrium lattice constant. A small expansion of the lattice leads
to the emergence of ferrimagnetism also in these compounds.

Although all compounds are not half-metallic at their equilibrium
lattice constant, small expansion of the lattice pushes the Fermi
level within the gap which is now situated in the majority-spin
band contrary to all other full-Heusler alloys. In table
\ref{tableHFerri} we have gathered the atom-resolved and total
spin moments for some of the compounds under study, for the
largest and smallest lattice parameters for which half-metallicity
is present. The total spin moment is $-2$ $\mu_B$ for the
Mn$_2$VAl(Ga or In) compounds which have 22 valence electrons per
unit cell and $-1$ $\mu_B$ for the Mn$_2$VSi(Ge or Sn) compounds
with 23 valence electrons (the negative sign of $M_t$ is according
to the convention set by equation~\ref{eq:SP2}). Thus these
compounds follow the Slater-Pauling behaviour and the ``rule of
24''. The lighter the element and the smaller the number of
valence electrons, the wider is the gap and the more stable is the
half-metallicity with respect to the variation of the lattice
constant.

\section{Effect of temperature on the spin polarization \label{sec11}}

The discussion so-far has been focused on the magnetic ground state of
half-metallic Heusler alloys. At elevated temperatures $T$ the spin
polarization $P(E_F;T)$ must clearly be reduced, and at the Curie
temperature $T_C$ it must vanish together with the average
magnetization. An qualitative and quantitative understanding of this
effect is interesting from the point of view of basic science, but is
also required in order to predict the applicability of half-metals in
realistic devices.

A theoretical study of the temperature dependence of $P(E_F)$
requires knowledge of the response of the spectral function
(density of states) to an increase of the temperature. This is a
two-step process. Firstly, one needs to find the response of the
magnetic excitations to temperature. Secondly, the spectral
function must be found in the presence of the excitations. Both
steps are not trivial, thus suitable approximations are needed. To
our knowledge, there have been three different approaches to the
subject for the case of half-metallic ferromagnets: a
tight-binding approach \cite{Dowben}, an approach based on
constrained density-functional theory combined with the concept of
a disordered local moment \cite{LezaicTemp}, and an approach based
on dynamical mean-field theory (DMFT) \cite{Chioncel,Itoh}.

Skomski and Dowben \cite{Dowben} consider a tight-binding model of a
half-metallic ferromagnet. Magnetic excitations, in the form of
magnons, will cause a rotation of the spin quantization axis, in
general different at each atom $i$, given by a unitary transformation
$U(\theta_i,\phi_i)$. This causes a projection of spin-up states of
one atom to spin-down states of another. They arrive at a crude
estimate of this spin-mixing contribution to the spin-down DOS:
$n_{\downarrow}\simeq n_{\uparrow} (M_0 - M(T))/(M_0 + M(T))$, in
terms of the spin-up DOS $n_{\uparrow}$, the magnetization $M_0$ at
$T=0$, and at temperature $T$, $M(T)$. This gives an estimate for the
polarization $P(T)\simeq M(T)/M_0$. This picture is appealing because
of its simplicity, and should be correct when the magnetic excitations
are predominantly of the Heisenberg type and have a long wavelength,
i.e., at low temperatures. Then, the local spin quantization axis is
unchanged over large distances (of the order of a fraction of the
magnon wavelength), and, at each instant, the material can be
considered locally half-metallic; the reduction of $P$ comes about
from the angle between the local spin quantization axes and the global
axis corresponding to the average magnetic moment.

However, as is argued by Le\v{z}ai\'c et al.~\cite{LezaicTemp}, at
higher temperatures the wavelength of the magnetic excitations can
become smaller and affect the local electronic structure. The
hybridization between atomic wavefunctions is crucial for the gap, and
the hybridization strength can be altered if the local spin axes of
neighbouring atoms are different. Therefore, the spectral function
should be in principle re-calculated self-consistently in the presence
of the excitations. Since the characteristic times of magnetic
excitations are much longer than the electron hopping times, the
adiabatic approximation within density-functional theory can be
employed: an external constraint is used to force a non-collinear
state, representing the magnon; under this constraint, the electronic
structure is calculated self-consistently within density-functional
theory.

Such an approach was applied in reference~\cite{LezaicTemp}, with
the conclusion that a non-collinear configuration can affect the
hybridization strongly enough that the polarization collapses. In
particular, it was found for NiMnSb that, under the constraint of
a spin spiral, the gap does not close easily, but the
minority-spin $d$ bands around the gap move in such a way that
they cross the Fermi level. Because the $d$ bands have a high
spectral weight, $n_{\downarrow}(E_F)$ increases abruptly when
these reach $E_F$, and the polarization collapses. The effect was
quantified by using the disordered local moment approach. Within
this method, the magnetic state at $T>0$ is approximated by a
state of the form Ni(Mn$^{\uparrow}_{1-x}$Mn$^{\downarrow}_{x}Sb$,
i.e., the spin is flipped a fraction $x$ of the Mn atoms in a
random way, representing a disordered moment at the Mn site. For
$x=0$ the ferromagnetic ground state is recovered, while $=0.5$
corresponds to $T_C$. The electronic structure of this disordered
state was found within the CPA and the KKR Green function method.
In this way, $P$ can be calculated as a function of $M$, and, if
$M(T)$ is independentlu known, one can find $P(T)$. The results
show a collapse of the polarization at about 0.4$T_C$, which
corresponds to room temperature.

Chioncel et al.~\cite{Chioncel,Chioncel06} have considered the effect
of the dynamic electron-magnon interaction, resulting from electron
correlations. According to this picture, there exist excitations
within the (spin-down) half-metallic gap as superpositions of spin-up
electron states and spin-down electron states with
magnons~\cite{Irkhin02}. These excitations, also called
non-quasiparticle states, produce at $T=0$ a spectral weight at
energies just above $E_F$ (or just below $E_F$ if the gap is among the
majority-spin states). For $T>0$, however, spin-down spectral weight
appears also at $E_F$, and the polarization is reduced. This
correlation-induced effect depends on the value of the on-site coulomb
interaction $U$, and can be captured within the DMFT. For the Heusler
alloy FeMnSb, calculations within the LDA+DMFT reported in
\cite{Chioncel06} show that, at $T=300K$, $P\simeq 0.9$ if $U=2$~eV
and $P\simeq 0.6$ if $U=4$~eV.

Unfortunately, there are not many experimental results concerning
the temperature dependence of the polarization in Heusler alloys.
One indirect result is provided in reference~\cite{Reiss04}, where
tunneling magnetoresistance (TMR) data are compared from junctions
containing a Co$_2$MnSi Heusler alloy as a first electrode with
data from junctions with Co$_{70}$Fe$_{30}$ or Ni$_{80}$Fe$_{20}$
(the second electrode is in all cases Co$_{70}$Fe$_{30}$, and the
barrier is AlO$_x$). It is shown that, at low temperatures, the
Co$_2$MnSi-based junction exhibits an 85\% TMR ratio (relative
change of resistance due to a change of alignment of the magnetic
moments of the leads), while the Co$_{70}$Fe$_{30}$- and
Ni$_{80}$Fe$_{20}$-based junctions exhibit 65\% and 70\% TMR ratio
respectively. At higher temperatures (above 150~K) the situation
is reversed, with the Co$_2$MnSi-based junction showing a lower
TMR ratio (30\% at 300~K) than the other two (45\% at 300~K). This
evidence suggests that, at low $T$, Co$_2$MnSi has a higher spin
polarization than Co$_{70}$Fe$_{30}$ or Ni$_{80}$Fe$_{20}$, but at
high $T$, the polarization of the Heusler alloy drops more
rapidly. We note here that Co$_2$MnSi has a high Curie temperature
of 985~K.

More recently, Sakuraba and collaborators \cite{Sakuraba} managed
to grow a Co$_2$MnSi/Al-O/Co$_2$MnSi tunnel junction. At low
temperatures (2~K), an extremely high TMR ratio of 570\% was
reported, while with increasing temperature the TMR dropped
rapidly, reaching about 67\% at 300~K. On the other hand, a
Co$_2$MnSi/Al-O/Co$_{75}$Fe$_{25}$ junction showed a less
pronounced decrease of TMR from 160\% at 2~K to 70\% at 300~K,
while a junction Co$_{75}$Fe$_{25}$/Al-O/Co$_{75}$Fe$_{25}$ showed
a less drastic decrease of TMR, approximately from 60\% to 30\%.
The authors of reference~\cite{Sakuraba} propose an explanation in
terms of spin-flip processes due to magnetic impurities at the
interface Co$_2$MnSi/Al-O. Alternatively, it is possible that
there is a rapid loss of half-metallicity with increasing
temperature, causing a decrease of $P$ much faster than $M$.

\section{Summary and Outlook} \label{sec12}

In this review we have given an introduction to the electronic
structure and the resulting magnetic properties of half-metallic
Heusler alloys, which represent interesting hybrids between
metallic ferromagnets and semiconductors.

First, we reviewed a few recent results on the electronic
properties of these alloys. The origin of the gap in these
half-metallic alloys and its connection to the magnetic properties
are well understood. The gap in half-Heuslers arises between the
bonding and antibonding $d$-hybrids created by the transition
metal atoms (e.g. Ni and Mn in NiMnSb). In full-Heusler compounds,
like Co$_2$MnSi, there is the extra complication of the appearance
of Co $d$-states which do not hybridize with the Mn states. There
are exactly 9 occupied minority-spin states for the half-Heuslers
and 12 for the full-Heuslers; thus the total spin moment follows a
Slater-Pauling behaviour for both families. Moderate changes of
the lattice parameter shift the Fermi level, but overall do not
affect half-metallicity. Spin-orbit coupling also induces states
within the gap but the alloys keep a very high degree of
spin-polarization at the Fermi level. Orbital moments are small as
expected for typical intermetallic ferromagnets.

We discussed the effect of substitutional doping, disorder and
defects on the properties of these alloys. In many cases moderate
degrees of disorder and doping do not alter half-metallicity. Also
defects with low formation energies, which are the most likely to
occur, keep the half-metallic character of the compounds.
Quaternary Heusler alloys are a special case of full-Heusler
alloys where one site is occupied randomly by two chemical
elements. These alloys, when the two extreme normal compounds are
half-metallic, follow the Slater-Pauling behaviour keeping the
half-metallicity of the normal Heusler alloys. Finally we discuss
the appearance of half-metallic ferrimagnetism in Mn$_2$VZ (Z and
$sp$ element).

The results summarized above  contribute in the idea that
half-metallic ferromagnets are feasible experimentally since all
studied effects keep an almost 100\% degree of spin-polarization
at the Fermi. However as discussed in the last section temperature
effects quickly destroy the half-metallicity and the alloys
transit to a normal metal. Moreover, surfaces/interfaces of
Heusler alloys are known from first principles calculations to
show strong surface/interface states which can severely reduce the
degree of spin-polarization \cite{GalaSurfInterf,Jenkins}. A
theoretical analysis~\cite{Mavropoulos05} shows that interface
states at a half-metal/semiconductor interface can reduce the
tunneling magnetoresistance ratio significantly.

\ack{The authors would like to thank their collaborators in the
research projects presented here: P.H. Dederichs, N. Papanikolaou,
R. Zeller, H. Ebert, V.  Popescu, K. \"Ozdo\~gan, E. \c Sa\c s\i
o\~glu, B. Akta\c s, L.M.  Sandratskii, P. Bruno, M. Le\v{z}ai\'{c},
G. Bihlmayer, and S.  Bl\"ugel.}

\end{document}